\documentclass[aps,prd,10pt,twocolumn,superscriptaddress,showpacs,nobibnotes,longbibliography]{revtex4-1}

\usepackage{gensymb}
\usepackage{hyperref}
\usepackage{graphicx}
\usepackage{amsfonts,amsmath,amssymb,bm,bbm}
\usepackage{color,soul}
\usepackage{slashed}
\usepackage[toc,page]{appendix}
\usepackage{lineno}
\usepackage{csquotes}

\hypersetup{
    pdfnewwindow=true,      
    colorlinks=true,       
    linkcolor=blue,          
    citecolor=blue,        
    filecolor=blue,      
    urlcolor=blue        
}

\begin{document}

\title{Limits on Sub-GeV dark matter from the PROSPECT reactor antineutrino experiment}

\affiliation{Brookhaven National Laboratory, Upton, New York, USA}
\affiliation{Department of Physics, Drexel University, Philadelphia, Pennsylvania, USA}
\affiliation{George W.\,Woodruff School of Mechanical Engineering, Georgia Institute of Technology, Atlanta, Georgia, USA}
\affiliation{Department of Physics \& Astronomy, University of Hawaii, Honolulu, Hawaii, USA}
\affiliation{Department of Physics, Illinois Institute of Technology, Chicago, Illinois, USA}
\affiliation{Nuclear and Chemical Sciences Division, Lawrence Livermore National Laboratory, Livermore, California, USA}
\affiliation{Department of Physics, Le Moyne College, Syracuse, New York, USA}
\affiliation{National Institute of Standards and Technology, Gaithersburg, Maryland, USA}
\affiliation{High Flux Isotope Reactor, Oak Ridge National Laboratory, Oak Ridge, Tennessee, USA}
\affiliation{Physics Division, Oak Ridge National Laboratory, Oak Ridge, Tennessee, USA}
\affiliation{Department of Physics, Temple University, Philadelphia, Pennsylvania, USA}
\affiliation{Department of Physics and Astronomy, University of Tennessee, Knoxville, Tennessee, USA}
\affiliation{Institute for Quantum Computing and Department of Physics and Astronomy, University of Waterloo, Waterloo, Ontario, Canada}
\affiliation{Department of Physics, University of Wisconsin, Madison, Wisconsin, USA}
\affiliation{Wright Laboratory, Department of Physics, Yale University, New Haven, Connecticut, USA}
\author{M.\,Andriamirado}
\affiliation{Department of Physics, Illinois Institute of Technology, Chicago, Illinois, USA}
\author{A.\,B.\,Balantekin}
\affiliation{Department of Physics, University of Wisconsin, Madison, Wisconsin, USA}
\author{H.\,R.\,Band}
\affiliation{Wright Laboratory, Department of Physics, Yale University, New Haven, Connecticut, USA}
\author{C.\,D.\,Bass}
\affiliation{Department of Physics, Le Moyne College, Syracuse, New York, USA}
\author{D.\,E.\,Bergeron}
\affiliation{National Institute of Standards and Technology, Gaithersburg, Maryland, USA}
\author{N.\,S.\,Bowden}
\affiliation{Nuclear and Chemical Sciences Division, Lawrence Livermore National Laboratory, Livermore, California, USA}
\author{C.\,D.\,Bryan}
\affiliation{High Flux Isotope Reactor, Oak Ridge National Laboratory, Oak Ridge, Tennessee, USA}
\author{T.\,Classen}
\affiliation{Nuclear and Chemical Sciences Division, Lawrence Livermore National Laboratory, Livermore, California, USA}
\author{A.\,J.\,Conant}
\affiliation{George W.\,Woodruff School of Mechanical Engineering, Georgia Institute of Technology, Atlanta, Georgia, USA}
\author{G.\,Deichert}
\affiliation{High Flux Isotope Reactor, Oak Ridge National Laboratory, Oak Ridge, Tennessee, USA}
\author{M.\,V.\,Diwan}
\affiliation{Brookhaven National Laboratory, Upton, New York, USA}
\author{M.\,J.\,Dolinski}\affiliation{Department of Physics, Drexel University, Philadelphia, Pennsylvania, USA}
\author{A.\,Erickson}
\affiliation{George W.\,Woodruff School of Mechanical Engineering, Georgia Institute of Technology, Atlanta, Georgia, USA}
\author{B.\,T.\,Foust}
\affiliation{Wright Laboratory, Department of Physics, Yale University, New Haven, Connecticut, USA}
\author{J.\,K.\,Gaison}
\affiliation{Wright Laboratory, Department of Physics, Yale University, New Haven, Connecticut, USA}
\author{A.\,Galindo-Uribarri}\affiliation{Physics Division, Oak Ridge National Laboratory, Oak Ridge, Tennessee, USA} \affiliation{Department of Physics and Astronomy, University of Tennessee, Knoxville, Tennessee, USA}
\author{C.\,E.\,Gilbert}\affiliation{Physics Division, Oak Ridge National Laboratory, Oak Ridge, Tennessee, USA} \affiliation{Department of Physics and Astronomy, University of Tennessee, Knoxville, Tennessee, USA}
\author{S.\,Hans}\affiliation{Brookhaven National Laboratory, Upton, New York, USA}
\author{A.\,B.\,Hansell}
\affiliation{Department of Physics, Temple University, Philadelphia, Pennsylvania, USA}
\author{K.\,M.\,Heeger}
\affiliation{Wright Laboratory, Department of Physics, Yale University, New Haven, Connecticut, USA}
\author{B.\,Heffron}\affiliation{Physics Division, Oak Ridge National Laboratory, Oak Ridge, Tennessee, USA} \affiliation{Department of Physics and Astronomy, University of Tennessee, Knoxville, Tennessee, USA}
\author{D.\,E.\,Jaffe}
\affiliation{Brookhaven National Laboratory, Upton, New York, USA}
\author{S.\,Jayakumar}\affiliation{Department of Physics, Drexel University, Philadelphia, Pennsylvania, USA}
\author{X.\,Ji}
\affiliation{Brookhaven National Laboratory, Upton, New York, USA}
\author{D.\,C.\,Jones}
\affiliation{Department of Physics, Temple University, Philadelphia, Pennsylvania, USA}
\author{J. Koblanski}\affiliation{Department of Physics \& Astronomy, University of Hawaii, Honolulu, Hawaii, USA}
\author{O.\,Kyzylova}\affiliation{Department of Physics, Drexel University, Philadelphia, Pennsylvania, USA}
\author{C.\,E.\,Lane}\affiliation{Department of Physics, Drexel University, Philadelphia, Pennsylvania, USA}
\author{T.\,J.\,Langford}
\affiliation{Wright Laboratory, Department of Physics, Yale University, New Haven, Connecticut, USA}
\author{J.\,LaRosa}
\affiliation{National Institute of Standards and Technology, Gaithersburg, Maryland, USA}
\author{B.\,R.\,Littlejohn}
\affiliation{Department of Physics, Illinois Institute of Technology, Chicago, Illinois, USA}
\author{X.\,Lu}\affiliation{Physics Division, Oak Ridge National Laboratory, Oak Ridge, Tennessee, USA} \affiliation{Department of Physics and Astronomy, University of Tennessee, Knoxville, Tennessee, USA}
\author{J.\,Maricic}\affiliation{Department of Physics \& Astronomy, University of Hawaii, Honolulu, Hawaii, USA}
\author{M.\,P.\,Mendenhall}\affiliation{Nuclear and Chemical Sciences Division, Lawrence Livermore National Laboratory, Livermore, California, USA}
\author{A.\,M.\,Meyer}\affiliation{Department of Physics \& Astronomy, University of Hawaii, Honolulu, Hawaii, USA}
\author{R.\,Milincic}\affiliation{Department of Physics \& Astronomy, University of Hawaii, Honolulu, Hawaii, USA}
\author{P.\,E.\,Mueller}\affiliation{Physics Division, Oak Ridge National Laboratory, Oak Ridge, Tennessee, USA} 
\author{H.\,P.\,Mumm}
\affiliation{National Institute of Standards and Technology, Gaithersburg, Maryland, USA}
\author{J.\,Napolitano}
\affiliation{Department of Physics, Temple University, Philadelphia, Pennsylvania, USA}
\author{R.\,Neilson}\affiliation{Department of Physics, Drexel University, Philadelphia, Pennsylvania, USA}
\author{J.\,A.\,Nikkel}
\affiliation{Wright Laboratory, Department of Physics, Yale University, New Haven, Connecticut, USA}
\author{S.\,Nour}
\affiliation{National Institute of Standards and Technology, Gaithersburg, Maryland, USA}
\author{J.\,L.\,Palomino}
\affiliation{Department of Physics, Illinois Institute of Technology, Chicago, Illinois, USA}
\author{D.\,A.\,Pushin}\affiliation{Institute for Quantum Computing and Department of Physics and Astronomy, University of Waterloo, Waterloo, Ontario, Canada}
\author{X.\,Qian}
\affiliation{Brookhaven National Laboratory, Upton, New York, USA}
\author{R.\,Rosero}
\affiliation{Brookhaven National Laboratory, Upton, New York, USA}
\author{P.\,T.\,Surukuchi}
\affiliation{Wright Laboratory, Department of Physics, Yale University, New Haven, Connecticut, USA}
\author{M.\,A.\,Tyra}
\affiliation{National Institute of Standards and Technology, Gaithersburg, Maryland, USA}
\author{R.\,L.\,Varner}\affiliation{Physics Division, Oak Ridge National Laboratory, Oak Ridge, Tennessee, USA} 
\author{D.\,Venegas-Vargas}\affiliation{Physics Division, Oak Ridge National Laboratory, Oak Ridge, Tennessee, USA} \affiliation{Department of Physics and Astronomy, University of Tennessee, Knoxville, Tennessee, USA}
\author{P.\,B.\,Weatherly}\affiliation{Department of Physics, Drexel University, Philadelphia, Pennsylvania, USA}
\author{C.\,White}
\affiliation{Department of Physics, Illinois Institute of Technology, Chicago, Illinois, USA}
\author{J.\,Wilhelmi}
\affiliation{Wright Laboratory, Department of Physics, Yale University, New Haven, Connecticut, USA}
\author{A.\,Woolverton}\affiliation{Institute for Quantum Computing and Department of Physics and Astronomy, University of Waterloo, Waterloo, Ontario, Canada}
\author{M.\,Yeh}
\affiliation{Brookhaven National Laboratory, Upton, New York, USA}
\author{C.\,Zhang}
\affiliation{Brookhaven National Laboratory, Upton, New York, USA}
\author{X.\,Zhang}
\affiliation{Nuclear and Chemical Sciences Division, Lawrence Livermore National Laboratory, Livermore, California, USA}

\collaboration{PROSPECT Collaboration}
\email{prospect.collaboration@gmail.com}

\author{Christopher V. Cappiello} 

\email{cappiello.7@osu.edu}

\affiliation{Center for Cosmology and AstroParticle Physics (CCAPP), Ohio State University, Columbus, Ohio 43210}

\affiliation{Department of Physics, Ohio State University, Columbus, Ohio 43210}

\date{\today}
\begin{abstract}
If dark matter has mass lower than around 1 GeV, it will not impart enough energy to cause detectable nuclear recoils in many direct-detection experiments. However, if dark matter is upscattered to high energy by collisions with cosmic rays, it may be detectable in both direct-detection experiments and neutrino experiments. We report the results of a dedicated search for boosted dark matter upscattered by cosmic rays, using $\sim$14.6 solar days of data from the PROSPECT reactor antineutrino experiment. We show that such a flux of upscattered dark matter would display characteristic diurnal sidereal modulation, and use this to set new experimental constraints on sub-GeV dark matter exhibiting large interaction cross sections.  

\end{abstract}

\maketitle


\section{Introduction}

Despite strong evidence for dark matter's (DM) existence, its particle nature remains unknown, and its identification is one of the most pressing problems in particle physics and astrophysics \cite{Ber04,Pet12,Ber16}. Direct searches for DM, focusing primarily on GeV-scale weakly interacting massive particles colliding with nuclei, are probing ever lower cross sections. However, these searches rapidly lose sensitivity for masses below about 1~GeV: if DM is too light, it does not impart enough momentum to trigger typical detectors. In recent years, a wide range of approaches has been explored in order to probe sub-GeV DM \cite{deN12,Iza13,Ali15,DeN16,Dol17,Ess17,Ibe17,Agn18D,Ake18,Glu18,Sla18,Xu18,Abr19,Arm19,Bel19,Cap19,Liu19,Nad19,Aga14,Nec16,Emk17,Giu17,Kac17,An18,Bri18,Ema18,Kim18,Yin:2018yjn,Alv19,Arg19,Bon19,Cap19b,Den19,Krn19,Abi20,Ema:2020ulo}. These include numerous studies of boosted DM, in which light DM is accelerated to high energy through a variety of processes~ \cite{Aga14,Nec16,Emk17,Giu17,Kac17,An18,Bri18,Ema18,Kim18,Yin:2018yjn,Alv19,Arg19,Bon19,Cap19b,Den19,Krn19,Abi20,Ema:2020ulo}.

One such process that would make light DM detectable is upscattering by cosmic rays (CRs). References ~\cite{Bri18,Ema18,Cap19b,Bon19,Den19,Ema:2020ulo} have explored the experimental signatures of DM particles being struck by CRs, upscattering to high energy, and interacting in direct-detection and/or neutrino experiments. Such analyses have the additional advantage of being insensitive to the cosmogenic DM velocity distribution, a source of uncertainty for traditional direct-detection limits that has received much attention recently \cite{Lis11, Mao14, Boz16, Slo16, Gre17, Nec18, Nec18b, Bes19, OHa19}. 
These analyses have constrained a wide range of parameter space for sub-GeV dark matter, but none of the experiments in question have performed their own analyses aimed at CR-upscattered DM.  
Using the PROSPECT reactor antineutrino detector, which combines the advantageous features of an on-surface deployment location with excellent particle discrimination capabilities, we have performed the first dedicated experimental analysis constraining sub-GeV DM by considering upscattering by CRs.  
This analysis, which is also the first to exploit the diurnal sidereal modulation of the boosted DM signal~\cite{Kouvaris:2014lpa,Ge:2020yuf}, addresses regions of  parameter space never before probed by terrestrial experiments.  
While cosmological limits do exist in this parameter space, they are model dependent and thus have faced some debate.  
Complementary constraints are valuable: while cosmological observables indirectly probe DM scattering in the early Universe, our analysis is based on scattering in the present day, making our analysis more comparable to traditional direct-detection studies than are cosmological limits.

This paper is organized as follows. In Sec.~\ref{sec:crdm}, we compute the flux of upscattered DM at Earth. In Sec.~\ref{sec:detector}, we describe the PROSPECT detector and the experimental search for DM. In Sec.~\ref{sec:analysis}, we discuss how we analyze data from PROSPECT to set limits on DM parameter space. In Sec.~\ref{sec:conclusions}, we present our limits.

\section{Upscattered Dark Matter Flux}\label{sec:crdm}

If the DM-nucleon scattering cross section $\sigma_{\chi N}$ is nonzero, CR nuclei have a chance to collide with DM particles as they propagate in the galaxy. In the DM mass range we consider in this paper, 1 keV $< m_{\chi} <$ 1 GeV, CR nuclei carry orders of magnitude more kinetic energy than galactic DM particles, and could upscatter DM particles to high energy. In this section, we compute the spectrum of energetic DM particles reaching Earth after being upscattered by CRs.

\subsection{Cosmic ray and dark matter inputs}

Our analysis depends on the Galaxy's DM density profile, the size of the CR halo, and the spectrum of CRs in the Galaxy. We consider only helium and proton CRs, as heavier nuclei are a small fraction of the flux, and including them would only marginally strengthen our limits. 

For the DM density, we assume a Navarro-Frenk-White (NFW) profile \cite{Nav96} with scale radius of $r_s = 20$ kpc (6.17$\times$10$^{20}$~m) and a density at Earth of 0.3 GeV/cm$^3$. Below, we denote the density $\rho_{\chi}(r,\theta,\phi)$ in spherical coordinates centered at Earth, i.e. $r$ is the distance from Earth and $\theta$ and $\phi$ denote the direction. As we show below, our results are not strongly affected by the differences between a NFW profile and a more shallow or cored profile near the Galactic Center.

Because we consider CRs scattering with DM throughout the Galaxy, we cannot naively employ the CR spectrum measured at Earth. The solar magnetic field suppresses the flux of low-energy CRs reaching Earth, contributing to the break in the observed spectrum at around 1 GeV/nucleon. Instead, we use the local interstellar spectrum (LIS) computed for protons and helium nuclei in Ref.~\cite{Bos17}. We assume that the energy distribution of CRs is independent of position in the CR halo, as the shape of the LIS has been shown by gamma-ray observations to be similar to the CR spectrum in other parts of the Galaxy \cite{fermi2011,Yang:2016jda}.

Galactic magnetic fields prevent CRs from simply streaming out of the Galaxy, binding them in a halo that is often modeled as a cylinder with a half-height of a few kpc. The exact value of the half-height is debated, but in this work we adopt a fairly standard value of 4 kpc (see Ref.~\cite{Str07} and references therein), and assume the CR density is independent of height within this halo and zero outside of it. The radial distribution, meanwhile, can be inferred from gamma-ray observations \cite{Rec16,Cat19}. We use for the radial profile the shallowest curve in Fig. 1 of Ref.~\cite{Cat19}, as it provides the best fit to relevant astrophysical gamma-ray observations. We neglect anisotropies in the CR flux \cite{Abeysekara:2018qho}, as they are small compared to statistical uncertainties in the data used (see below). We assume the CR density is zero beyond a cylinder radius of 25 kpc. Thus, our model for the CR density is

\begin{equation}
\rho_{\rm CR}(r_{cr},z) = \rho_0g(r_{cr})\Theta(25 - r_{cr})\Theta(4 - \left|z\right|)\,,
\end{equation}
where $g(r_{cr})$ is the radial distribution taken from Ref.~\cite{Cat19}, $\rho_0$ is a normalization density fit to the LIS at Earth's position in the Galaxy, and $r_{cr}$ and $z$ are in units of kpc, with the origin at the Galactic Center. In our analysis, we do not consider the change in the Earth's position over the course of the data taking, as the amount the Earth moves in approximately two weeks is small compared to galactic distance scales.

\subsection{Cosmic ray-dark matter scattering}

As in Refs.~\cite{Bri18,Cap19,Cap19b}, we assume an energy-independent DM-nucleon cross section $\sigma_{\chi N}$. While the feasibility of producing a model with this scaling has been questioned in the literature \cite{Krn19}, its use enables useful comparisons to other direct-detection limits. Specific models involving light mediators have also been applied to CR-DM scattering \cite{Den19,Bon19}.  

Because CR velocities are so much higher than the velocity of DM bound in the Galaxy, we can treat the DM particles as being at rest. The kinetic energy $T_{\chi}$ transferred to a DM particle of mass $m_{\chi}$ by a CR of mass $m_{\rm CR}$ and kinetic energy $T_{\rm CR}$ is

\begin{equation}
T_{\chi} = \frac{T_{\rm CR}^2 + 2m_{\rm CR}T_{\rm CR}}{T_{\rm CR} + (m_{\rm CR} + m_{\chi})^2/(2m_{\chi})}\left(\frac{1-\cos \theta}{2}\right)\,,
\end{equation}
where $\theta$ is the center-of-mass scattering angle \cite{Cap19b}. We denote the maximum kinetic energy transferred to a DM particle $T_{\chi}^{\rm max}$:

\begin{equation}
T_{\chi}^{\rm max} = \frac{T_{\rm CR}^2 + 2m_{\rm CR}T_{\rm CR}}{T_{\rm CR} + (m_{\rm CR} + m_{\chi})^2/(2m_{\chi})}\,.
\end{equation}

Given the LIS of CR species $i$, $d\Phi_i/dT_i$, and the DM density $\rho_{\chi}$, we can compute the spectrum of upscattered DM in terms of CR kinetic energy,

\begin{equation}\label{DMdist}
\frac{d\Phi_{\chi}}{dT_i}=\int\frac{d\Omega}{4\pi}\int_{l.o.s.}dl \, \sigma_{\chi i}\frac{\rho_{\chi}}{m_{\chi}}\frac{d\Phi_i}{dT_i}\,.
\end{equation}
The line-of-sight integral extends from Earth to the edge of the CR halo. From here we compute the spectrum in terms of DM kinetic energy $T_{\chi}$:

\begin{equation}\label{DMdist2}
\frac{d\Phi_{\chi}}{dT_{\chi}} = \int_0^{\infty}dT_i \frac{d\Phi_{\chi}}{dT_i} \frac{1}{T_{\chi}^{\rm max}(T_i)}\Theta[T_{\chi}^{\rm max}(T_i)-T_{\chi}]\,.
\end{equation}
We consider DM boosted to energies between 25 MeV and 1 GeV. The low-energy limit is determined by the analysis threshold (described below), while the high-energy cutoff allows us to neglect quasielastic and inelastic processes \cite{Bri18}. For more discussion and a visualization of the upscattered DM spectrum, see Ref.~\cite{Cap19b}.

In this analysis, we assume the standard cross section scaling for spin-independent DM-nucleon scattering. In the limit of zero momentum transfer, the DM-nucleon cross section $\sigma_{\chi N}$ is related to the DM-nucleus cross section $\sigma_{\chi A}$ (for a nucleus of mass number A) via the formula

\begin{equation} 
\sigma_{\chi A} = \left(\frac{\mu_{\chi A}}{\mu_{\chi N}}\right)^2 A^2 \sigma_{\chi N}
\end{equation}
where $\mu_{\chi N}$ and $\mu_{\chi A}$ are the DM-nucleon and DM-nucleus reduced masses, respectively. For nonzero momentum transfer, the differential cross section is modified by a form factor $F(q^2)$ as
 
\begin{equation}
\frac{d\sigma_{\chi A}}{dq^2} = \frac{d\sigma_{\chi A}}{dq^2}|_{q^2 = 0} |F(q^2)|^2. 
\end{equation}
For nuclei heavier than hydrogen, we use the standard Helm form factor \cite{Hel56,Dud06}, and for protons, we use the dipole form of the hadronic form factor from Ref.~\cite{Per06}, both in accordance with Refs.~\cite{Bri18,Cap19b}.

Although this upscattering could have been happening for billions of years, DM boosted to such high energies is no longer gravitationally bound within the galaxy. This prevents a long-term buildup of a high-energy DM in the galaxy, and the flux reaching Earth is relatively small. For DM with a mass of 1 MeV and scattering cross section of $10^{-30}$ cm$^2$, the DM flux reaching Earth is approximately $1.26 \times 10^{-6}$ cm$^{-2}$ s$^{-1}$ sr$^{-1}$.

\subsection{Propagation to the detector}

At the large cross sections we consider, DM may scatter many times while passing through the atmosphere and detector shielding. If the cross section is too large, DM may lose too much energy to be detectable or simply be scattered back out of the atmosphere. The exclusion regions derived in Refs.~\cite{Bri18,Cap19b,Den19,Bon19} all have ceilings, cross sections above which DM would be mostly or entirely blocked from triggering the detector in question. So it is critical to model DM scattering with nuclei as it travels through the atmosphere and detector shielding. 

For particles arriving from above the horizon, we simulate propagation through the atmosphere and detector shielding using the same propagation code used in Ref.~\cite{Cap19b}. This code generates $10^6$-$10^7$ DM particles at the top of the atmosphere, where each particle is assigned an incoming direction and initial kinetic energy. The kinetic energy is drawn from the spectrum of incoming DM, while the incoming direction is drawn from a distribution of incoming $\theta$ and $\phi$, which is in turn based on the direction dependence of the line of sight integral of DM density times CR density. Each particle is then propagated through the atmosphere.  
The distance to the first collision is drawn from a probability distribution based on the DM's mean free path, the scattering angle is drawn from an isotropic distribution in the center-of-mass frame, and the energy loss and lab-frame scattering angle is computed based on the incoming energy, scattering angle, and target nucleus.

This process is repeated until the particle reaches a sea-level modeled detector location, is scattered out of the atmosphere, or loses too much energy to be detected.  
The minimal ($<$1~m water equivalent) concrete overburden provided by the building surrounding the on-surface PROSPECT detector unsurprisingly plays a very minor (percent-level) role in attenuation and down scattering of the DM flux.  Reduction in atmospheric overburden due to PROSPECT's  $\sim260$~m elevation relative to sea level is similarly negligible.


At the cross sections we consider, any DM particles arriving from below the horizontal direction are completely blocked.  
This causes the flux at the detector to vary over the course of a day as the Earth's bulk rotates in front of PROSPECT and blocks fluxes from galactic locations with greater or lesser boosted DM density.  

The distribution of scattering angles during propagation, and the resulting proton recoil spectra, are determined based on the corresponding nuclear form factors. During propagation, the suppression of the total DM-nucleus cross section due to the form factor is neglected, a conservative choice that follows Ref.~\cite{Bri18}, and has the effect of reducing the flux of very high-energy DM which, as mentioned above, may suffer inelastic collisions. Including such suppression would not change the fact that DM is blocked by the Earth, and would only make our results sensitive to somewhat higher cross sections.

As the flux depends on the location of PROSPECT relative to celestial objects (rather than to the Sun), it is expected that these modulations would exhibit a period of one sidereal day (rather than one solar day).  
For a given DM mass, cross section, and time, we use the aforementioned propagation code to compute the DM flux at the location of the PROSPECT detector.  
As detailed below, this allows us to compute the predicted diurnal sidereal modulation of the detected DM signal in PROSPECT and compare it to any time dependence observed in the experiment's data.  

\section{Experimental Search}\label{sec:detector}
The PROSPECT experiment was designed to measure the energy spectrum of antineutrinos emitted from the highly $^{235}$U-enriched High Flux Isotope Reactor (HFIR) reactor at Oak Ridge National Laboratory~\cite{prospect}.  
Through simultaneous measurements at many $<$10~m reactor-detector distances, PROSPECT is able to probe the existence of short-baseline electron antineutrino disappearance caused by oscillation between active and sterile neutrino states~\cite{anomaly_white,vsbl} with greatest sensitivity in the 1-10~eV$^2$ scale mass splitting regime~\cite{prospect_osc,prospect_prd}.  
By integrating measurements over all baselines, PROSPECT also provides world-leading precision in measuring antineutrino production by $^{235}$U fission products~\cite{prospect_spec,prospect_prd}.  
Beyond reactor oscillation and spectrum physics, the PROSPECT detector's on-surface location and powerful particle identification capabilities provide a unique opportunity for DM detection.  

\subsection{Experiment and dataset description}
The PROSPECT detector consists of an 11$\times$14 array of 1.2~meter long optically isolated segments filled with $^6$Li-doped liquid scintillator~\cite{prospect_ls}.  
Each segment is equipped with a photomultiplier tube (PMT) at each end for collecting scintillation light produced by charged particle interactions.  
Thin, specularly reflecting segment walls~\cite{prospect_grid} efficiently direct scintillation light towards these PMTs with less than 50\% variation in absolute light collection at all points inside a segment.  
Heavy charged particles producing high ionization density in the PROSPECT scintillator, such as protons, generate a characteristically longer light emission profile than light charged particles, such as electrons, enabling powerful particle identification capabilities via pulse shape discrimination (PSD).  
The central PSD-capable scintillator detector is surrounded on all sides by tens of cm of passive gamma and neutron shielding.  
As the primary physics goals of the PROSPECT experiment necessitate it being close to the HIFR core, the detector was deployed on the Earth's surface inside the HFIR building at a location with $<$1~m water-equivalent overburden.  
A more detailed description of the PROSPECT experimental layout and detector is available in Ref.~\cite{prospect_nim}.  

The data analyzed for this paper were collected during $\sim$14.6~solar days of detector operation from March 16th to March 31st, 2018, and are a subset of the datasets used in Refs.~\cite{prospect_osc,prospect_prd}.  
The HFIR reactor was not operating during the entire data-taking period.  
The triggering of PROSPECT's waveform digitizer (WFD) readout, described in detail in Refs~\cite{prospect_nim,prospect_prd}, occurs when PMTs on one detector segment observe time-aligned waveform features above approximately five~photoelectrons in amplitude.  
To reduce data rates, only sections of digitized waveforms above two~photoelectrons in amplitude are recorded to the data stream.  
With these settings, raw trigger rates were $\sim$2000~s$^{-1}$, with a trigger threshold of $\sim$75~keV in electron-equivalent energy.  
The 14-bit WFDs, with implemented PMT and electronics gain settings, exhibit linear response below 14~MeV electron-equivalent energy, beyond which point electronics saturation results in clipped waveforms.  
During the data-taking period, one or two PMT channels on 28 of 154 segments had experienced PMT voltage divider instabilities and were turned off.  
For simplicity, all data from these 28 segments were not considered.  

\subsection{Reconstructed physics quantities}

PROSPECT low-level data processing, calibration, and physics metric reconstruction is discussed in detail in Ref.~\cite{prospect_prd}.  
In the present analysis, we take advantage of reconstructed time, segment number, position, energy, and pulse shape variables to select candidate DM interactions.   
These reconstructed variables are separately assigned to each time-aligned waveform pair in one segment, called a pulse.  
Time-aligned pulses from different segments are then grouped to form larger data objects, which are referred to as clusters.  
The reconstructed position of a pulse along a segment (z position) is formed from relative charge integral and arrival time offsets between waveforms; previous analysis has demonstrated a resolution of $\sim$5~cm or better on reconstructed z positions for pointlike energy depositions.  

The reconstructed electron-equivalent energy (MeV) of a pulse is formed using the combined charge information from both PMTs.  
The absolute scale of reconstructed energy is defined using radioactive gamma-ray calibration sources and naturally occurring radiogenic and cosmogenic beta-particle and gamma-ray backgrounds~\cite{prospect_prd}.  
Electron-equivalent energy depositions by heavier charged particles are modeled using Birks quenching parameters~\cite{Birks} fitted to pulse data from triton-alpha products of n-Li capture, radiogenic alpha-particle decays, and collaboration-external measurements of liquid scintillator proton quenching factors~\cite{quench3,quench2,quench1}.  
The accuracy of these external measurements in describing PROSPECT data was further validated using proton recoil spectra generated by deployment of a $^{241}$Am-$^9$Be fast neutron source inside the central detector.  

For pointlike energy depositions in the scintillator, reconstructed pulse energy resolution is dominated by a photo-statistics contribution of approximately (4.5/$\sqrt{E}$)\%; reconstructed energy scales are stable to $<$1\% in time and segment number.  
A reconstructed pulse shape variable is formed from the ratio of late ($>$44~ns after the leading pulse edge) to total charge (between 12~ns before and 200~ns after the leading pulse edge) in a pulse.  

\begin{figure}[hptb!]
	    \centering
   	    \includegraphics[width=\columnwidth]{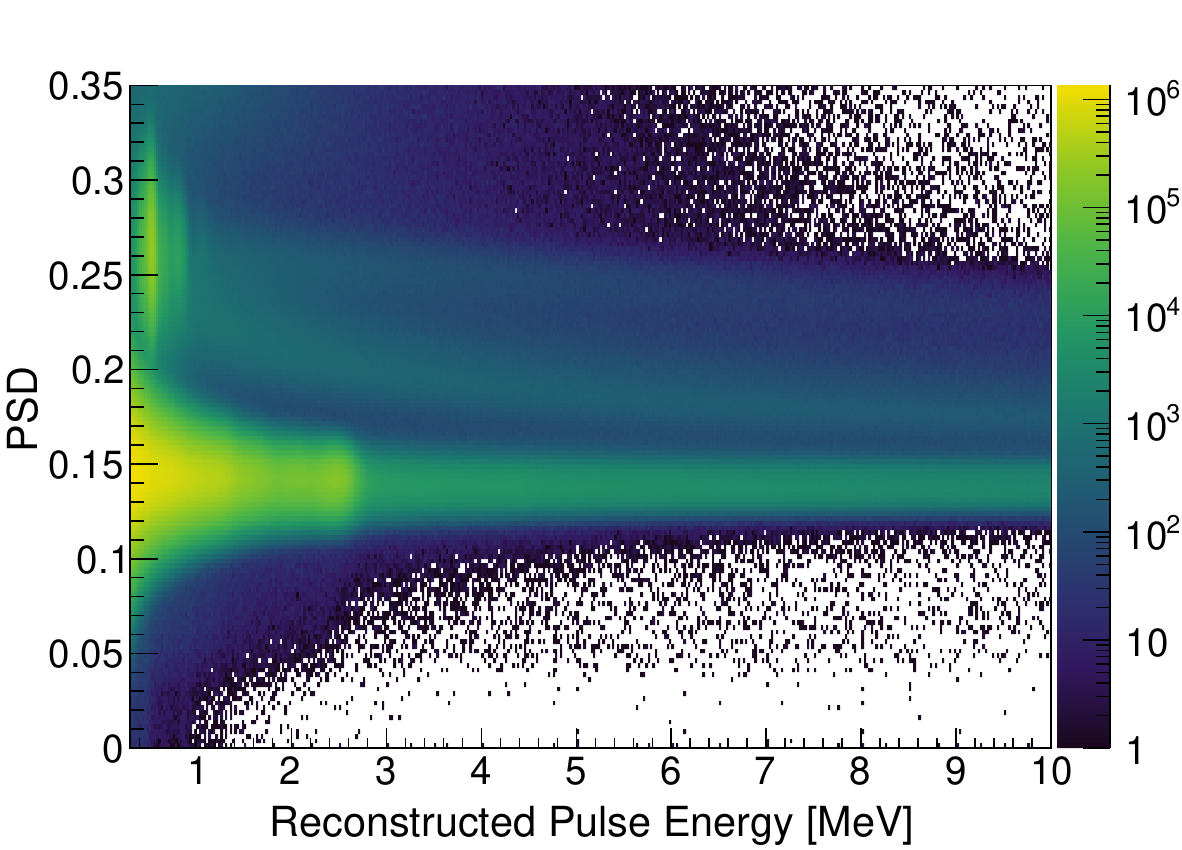}
	    \caption{Reconstructed PSD parameter values versus reconstructed energy for all single-pulse clusters in the full analysis dataset.}
	    \label{fig:PSDE}
	\end{figure}

For illustration, reconstructed energy versus PSD values for  a representative sample of single-pulse clusters are shown in Fig.~\ref{fig:PSDE}.  
Three distinct PSD bands are visible, corresponding to electron, proton, and nuclear recoil signatures from low to high PSD value.  
In the low PSD band, higher event rates at lower energy are clearly visible, including a prominent edge at 2.6~MeV contributed by ambient radiogenic $^{208}$Tl.  
Low-energy features are also visible in the higher-PSD bands, including a prominent mono-energetic peak from n-Li capture, as well as $<$1.25~MeV features from $^{215}$Po, $^{214}$Po, $^{212}$Po, and other alpha-particle decays.  
Higher energy events in all bands arise almost entirely from interactions of cosmic muons and neutrons with the detector.  
Event rates in the electronlike recoil band far outnumber those in the proton or nuclear recoil bands.  

\subsection{Signal selection and background reduction}

Using the variables above, we have selected detector signals consistent with DM scattering off of free protons (hydrogen nuclei) in the liquid scintillator.    
Scattered protons are expected to deposit all their energy in a single segment.  
For the scattering cross sections probed in this analysis, at most one interaction is expected per incident DM particle, with vertices evenly distributed throughout the detector active volume.  

\begin{figure}[hptb!]
	    \centering
	    \includegraphics[width=\columnwidth]{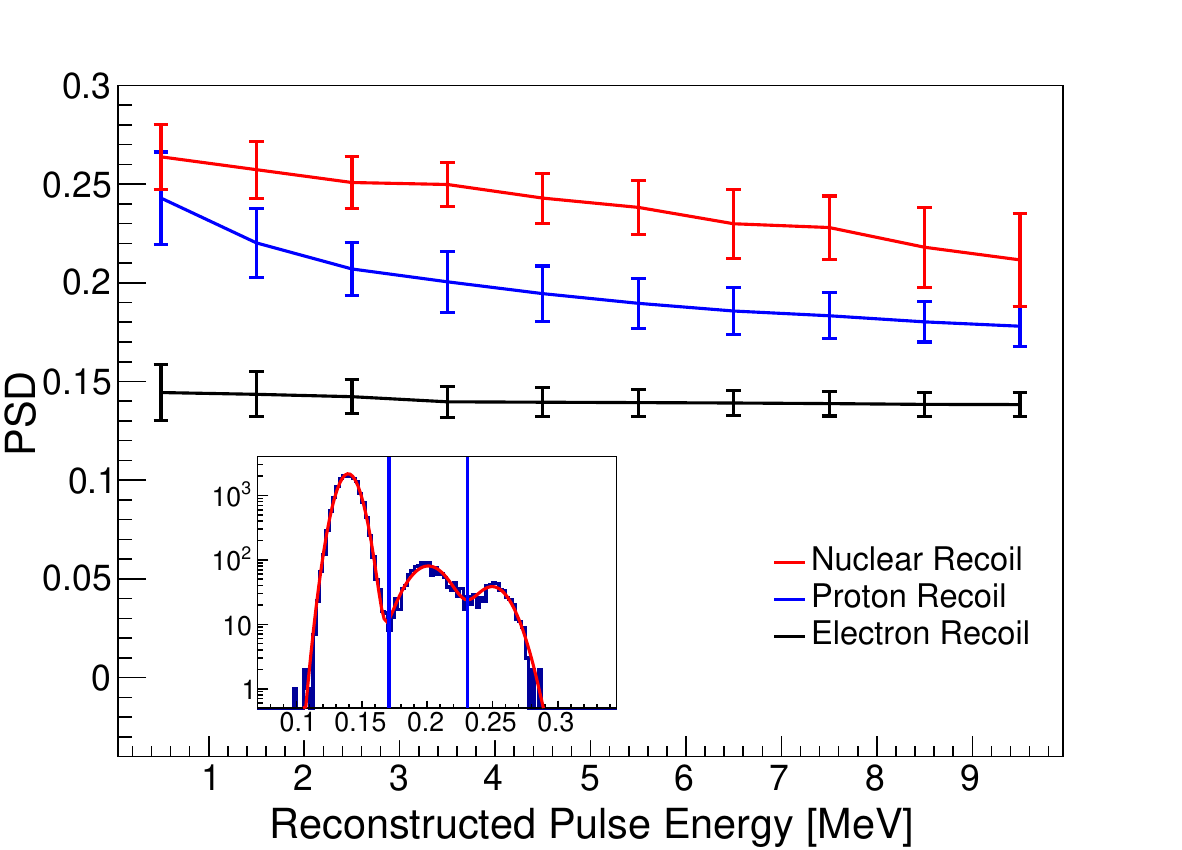}
	    \caption{Mean and 1$\sigma$ standard deviation of PSD parameter distributions for three particle types as a function of energy.  The inset image shows the underlying fitted distribution for the 3-4~MeV energy range, as well as blue lines representing the 2$\sigma$ width of the proton band in this energy range.}
	    \label{fig:PSDParameters}
	\end{figure}

As our signal definition, we select reconstructed clusters containing only one pulse, which must have a PSD value consistent with the protonlike recoil band shown in Fig.~\ref{fig:PSDE}.  
Allowed PSD ranges were defined by applying three-Gaussian fits to individual energy slices for all single-pulse clusters within 55~cm of the detector z center.   
Resulting fitted means and 1$\sigma$ widths are shown in Fig. ~\ref{fig:PSDParameters}.  
A representative fit for the 3.0-4.0 MeV energy range is shown in the figure inset.    
Candidate events are required to have a pulse PSD value within 2$\sigma$ of the proton recoil band mean, with exact cut values at each energy determined via linear interpolation between fitted values of adjacent energy bins.  
This cut carries a 5\% signal inefficiency, which is propagated through the analysis.  
PSD parameter distributions are sufficiently time stable as to expect negligible variation in signal efficiency or background contamination over the considered dataset.  
Due to the expected signal topology described above and the relatively low expected rate of accidentally coincident backgrounds, the single-pulse requirement has negligible associated inefficiency.  
The single-segment signal topology also ensures that inactive segments do not affect signal selection efficiency or energy response in functioning segments.  

As an on-surface detector, PROSPECT is subjected  to a high rate of incident cosmogenic neutrons and muons.  
Both of these particle types are capable of generating single-pulse, high-PSD protonlike signatures in PROSPECT, whether via scattering of primary neutrons from protons in the scintillator, or via scattering of secondary neutrons produced in nearby interactions of those primary particles.  
To reduce potential backgrounds from these sources, which are expected to dominate after application of all selection cuts, a series of fiducialization and time-coincidence cuts were applied to selected signal candidates.  
Due to their comparatively higher cross section, incident cosmic or secondary neutrons will preferentially produce proton recoil signatures on the outer edges of the liquid scintillator.  
Thus, all single-hit clusters reconstructed in segments in the two outermost rows and columns were rejected, as well as those reconstructed further than 20~cm from the z center of the detector.  
The third-to-bottom row of the detector was also removed from the selection due to higher trigger rates from the detector bottom primarily caused by imperfections in lead shielding coverage.  

Cosmogenic time-coincidence veto cuts were optimized to maintain high efficiency ($>$98\%) while still keeping cut lengths substantially longer than the associated physics timescale in question.  
To reject signals produced by scattering of secondary neutrons, all signals occurring within 5~$\mu$s of a preceding muonlike (energy greater than 15~MeV) cluster or 5~$\mu$s of a preceding or following proton recoil-like (containing at least one high PSD value pulse) cluster are rejected.  
All signals occurring less than 500~$\mu$s prior to a n-Li capture signal were also rejected.  
Descriptions of n-Li, neutronlike, and muonlike event class requirements are described in further detail in Ref.~\cite{prospect_prd}.  
These cosmogenic veto cuts each have an associated dead time of $<$1.5\%, which is corrected for in the analysis. 
Finally, to reject waveforms truncated by readout window boundaries as well as $^{212}$Po $\alpha$ particle signals in coincidence with preceding $^{212}$Bi~$\gamma+\beta$ decay signals (0.299~$\mu$s half-life), all signals occurring within 2~$\mu$s of any other trigger were rejected.
This `pileup' cut has $\sim$1\% associated dead time, which is also corrected for.  
Negligible ($<$0.1\%) variations in associated veto efficiencies are observed during the data-taking period.  

The rare event search performed in this paper rests on the assumption that any observed diurnal sidereal modulation in the rate of detected signal candidates arises from variations in the flux of DM traversing the PROSPECT detector.  
However, variations in signal-like event rates may also arise from  modulations in the flux of incident of cosmic neutrons and muons~\cite{nucifer,stereo_2019,prospect_prd}: a relative reduction in the flux of incident cosmic neutrons will result in similar reductions in neutron-proton recoils inside of PROSPECT, which are largely indistinguishable from DM-proton recoils.  
To quantify the level of expected cosmogenic variation in the two~week PROSPECT dataset, we use the n-Li capture dataset described above, which occurs at a high rate ($\sim$10~s$^{-1}$) in the detector.  
We note that PROSPECT has previously demonstrated that differing cosmogenically induced event types, including n-Li captures, show consistent rate fluctuations in response to changing atmospheric pressure~\cite{prospect_prd}.  
Average n-Li capture rates for each of the 24 sidereal hours in the sidereal day were divided by the total averaged n-Li capture rate to obtain hourly correction factors; these factors were then applied to various signal predictions used to perform the DM exclusion analysis described in the following section.  
Correction factors for each sidereal hour, depicted in the following section (Fig.~\ref{fig:asymmetry}), have an associated statistical uncertainty of 0.2\%, and are all within 1.5\% of unity.  


\subsection{Final candidate dataset and cross-checks}

\begin{figure}[hptb!]
	    \centering
	    \includegraphics[width=\columnwidth]{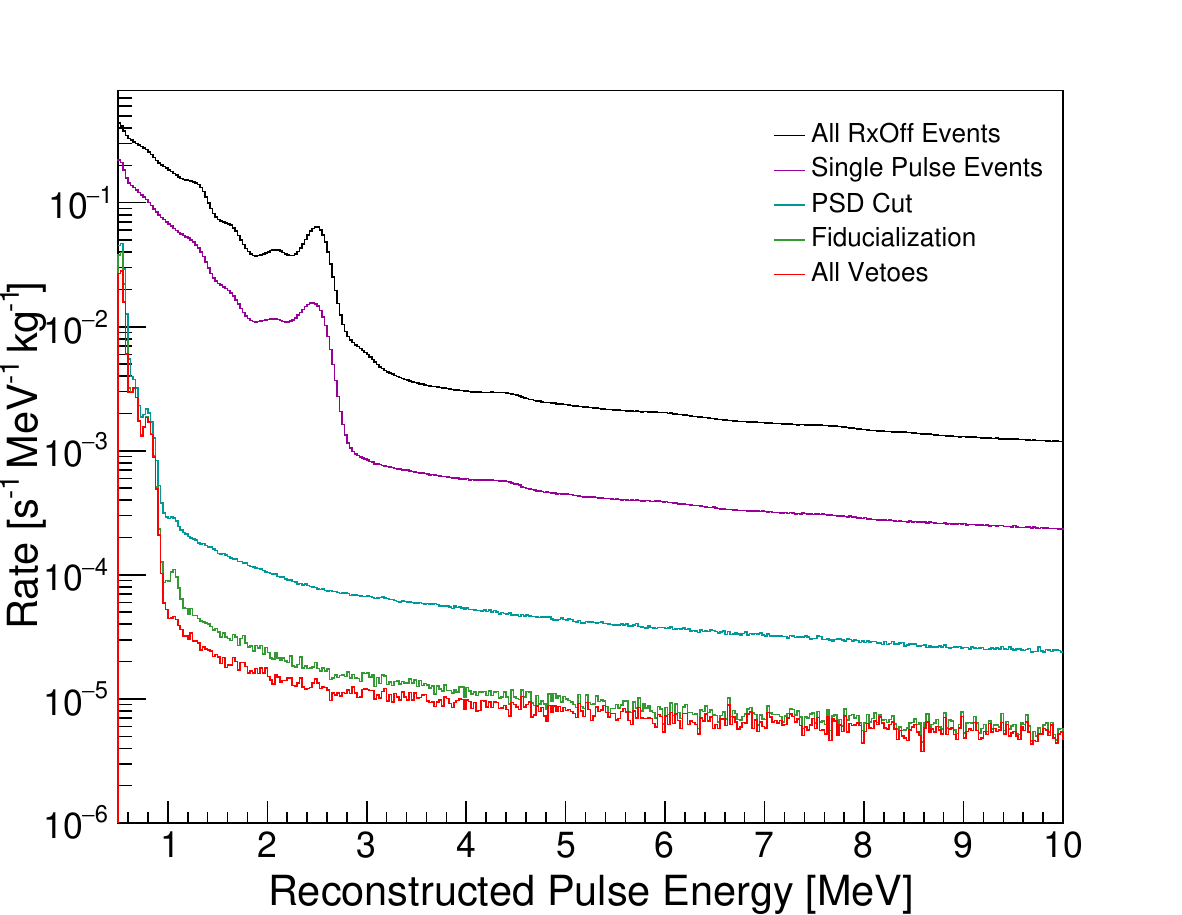}
	    \caption{Signal events obtained after sequential application of all selection cuts and background vetoes. The lowest red line represents the final  candidates used for further analysis.}
	    \label{fig:eventRate}
	\end{figure}

Efficiency-corrected signal count rates per~kg of scintillator obtained after applying all selection cuts are shown in Fig.~\ref{fig:eventRate}.  
Event rate reductions from initial proton-recoil-like criteria range from roughly 2 orders of magnitude at high energy to nearly 3 at lower energies, where the PSD requirement largely eliminates previously dominant ambient gamma-ray contributions.  
Subsequent background cuts contribute almost an additional order of magnitude reduction at most energies.  
After application of all cuts, DM-like event rates are as low as 5$\times$10$^{-6}$~s$^{-1}$\,MeV$^{-1}$\,kg$^{-1}$ at the highest considered energies in this analysis, or $\sim$150~d$^{-1}$\,MeV$^{-1}$ in the 440~kg fiducial detector volume.  
Below 1.5~MeV, signal rates begin to increase substantially.  
For this reason, in the DM analysis that follows, only events above 1.5~MeV are considered.  

Over the 14.6~solar day dataset, a total of 37522 DM-like signal candidates between 1.5 and 10~MeV (4.8 and 18.5 MeV$_{NR}$) are observed.  
It is expected that this candidate set is dominated by backgrounds consisting of a single proton recoil induced by a scattering cosmic neutron.  
The clear separation of PSD bands demonstrated in Fig.~\ref{fig:PSDParameters} ensures that electron recoil events, which are both cosmogenic and radiogenic in origin, are subdominant contributors to signal rates.  
Meanwhile, rates of cosmogenically induced nuclear recoils are lower than that of proton recoils, and are also expected to contribute subdominantly to the final signal dataset.

\begin{figure}[hptb!]
	    \centering
	    \includegraphics[width=\columnwidth]{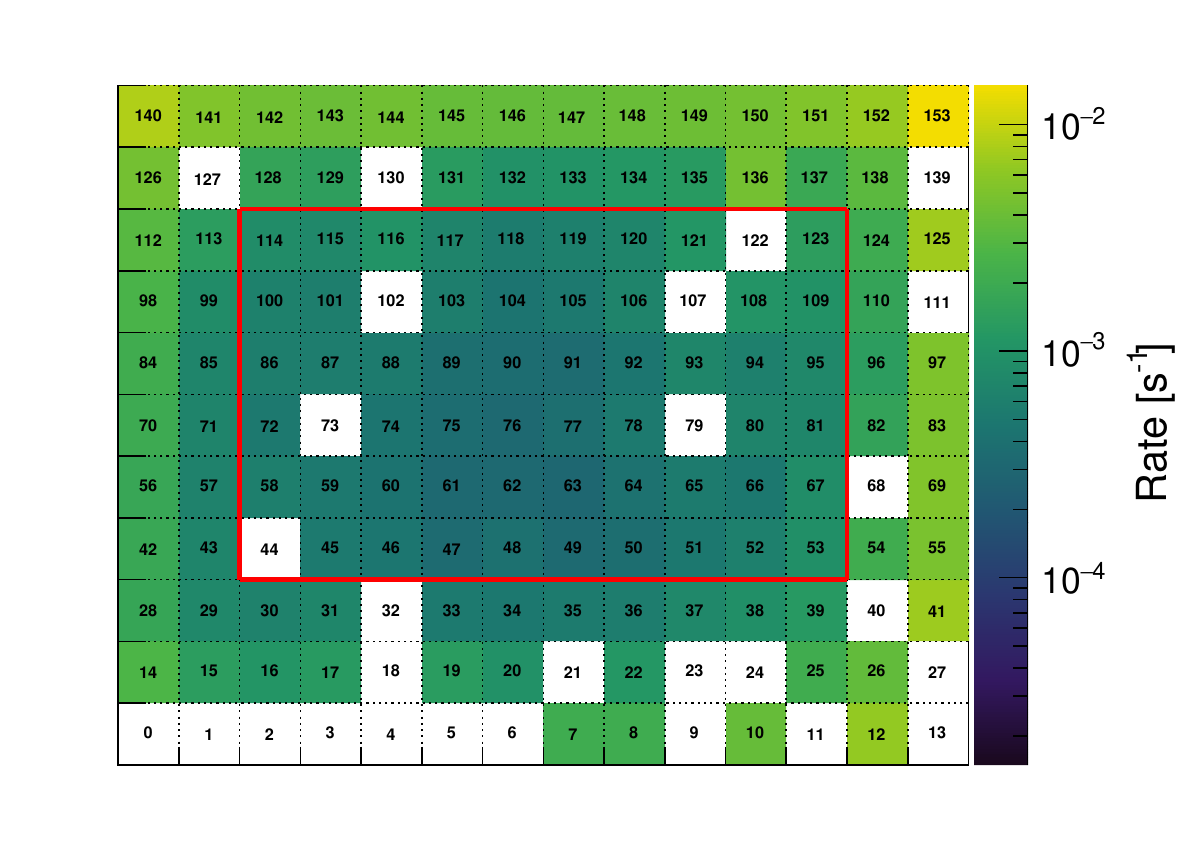}
	    \caption{Signal-like event rates per segment after application of all signal selection and background veto cuts.  Fiducial segments appear inside the red rectangle, while inactive segments are represented in white.  Higher segment numbers correspond to the top of the detector, while segments in the left-hand bottom corner are closest to the HFIR reactor core.}
	    \label{fig:eventRateSegment}
	\end{figure}

As shown in Fig.~\ref{fig:eventRateSegment}, signal events are relatively evenly distributed among the different fiducial detector segments, with per-segment signal rates of $\sim$0.5~ms$^{-1}$ between 1.5 and 10~MeV.  
This figure also illustrates the locations of inactive segments not used in the analysis, as well as the higher rates  of signal-like events in nonfiducial segments.  

To test for unforeseen variations in signal selection efficiency or improperly estimated cosmogenic flux variations, we compared populations of signal-like events occurring only in nonfiducial segments, since these signals, similar to those from the fiducial volume, are expected to be dominated by conventional cosmogenic neutron backgrounds.  
Two samples of roughly equal live time were formed from events with time stamps between the hours of either 22:00 to 02:00 or 10:00 to 14:00 Greenwich Mean Sidereal Time (GMST), which represent the periods of highest and lowest expected DM fluence through PROSPECT.  
These nonfiducial datasets contain $\sim$20 times higher statistics than their counterpart fiducial signal datasets.  

	\begin{figure}[hptb!]
	    \centering
	    \includegraphics[width=\columnwidth]{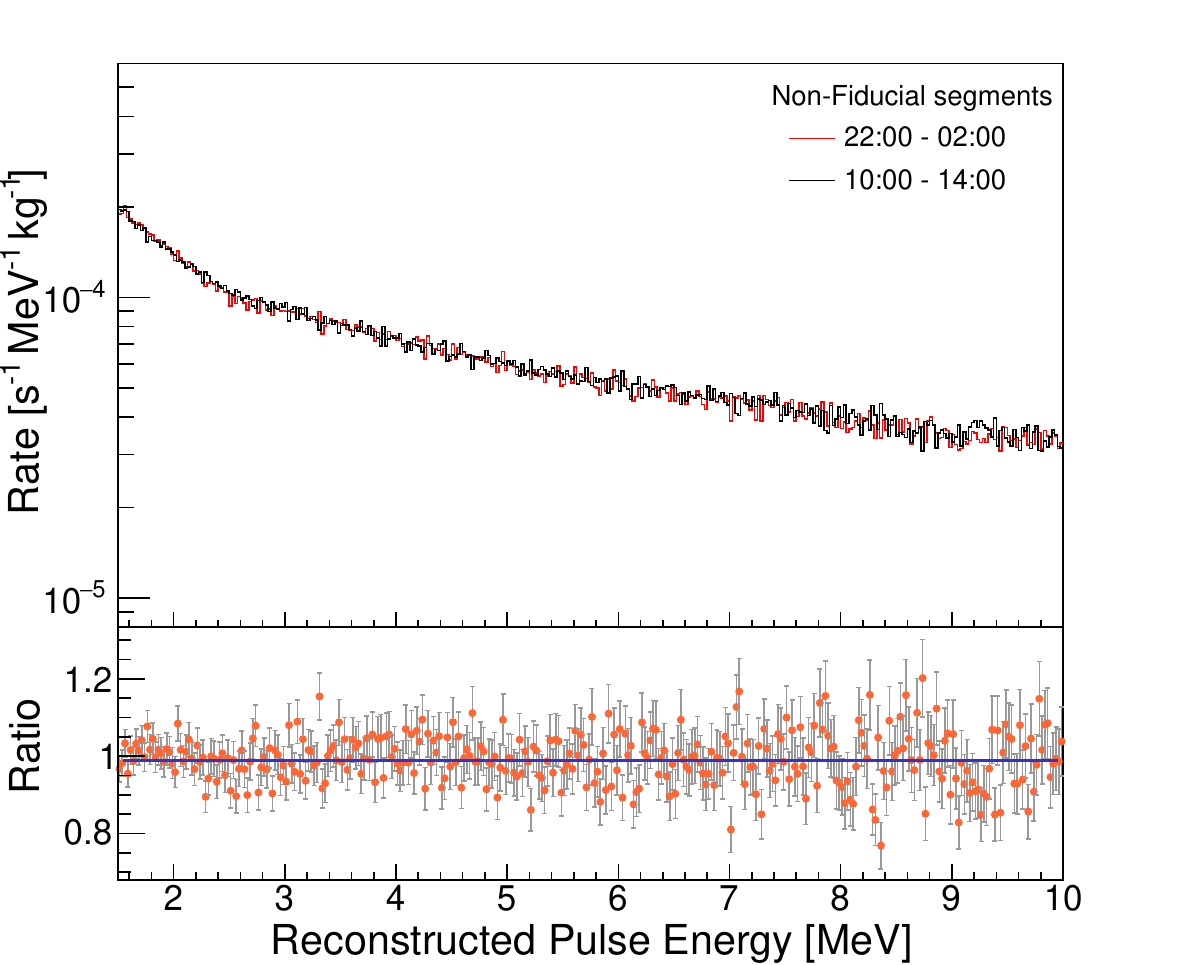}
	    \caption{Top: measured energy spectra for proton recoil-like events not selected for inclusion in the DM-like signal event sample due to their occurrence in non-fiducial PROSPECT detector segments.  Bottom: spectral ratio between these datasets.  The blue line represents a flat fit to this ratio, which is statistically consistent with unity.}
	    \label{fig:SignalStability}
	\end{figure}

The ratio of energy spectra between these two test data samples are plotted versus reconstructed energy in Fig.~\ref{fig:SignalStability}.  
Good consistency in event rates can be seen across the 1.5-10 MeV energy range of interest within the statistical limitations of the dataset.  
A flat-line fit to this ratio provides a best fit of 0.987$\pm$0.003, in agreement with the 0.988 value expected based on the hourly correction factors calculated above using the cosmic n-Li dataset.  
This agreement indicates no unexpected diurnal sidereal modulation in background rates within the statistical limitations of this comparatively large dataset.  
To test for possible modulation in energy spectrum shape, a linear polynomial was fit to the ratio between datasets.  
The best-fit slope parameter is within 1 standard deviation of zero, as would be expected from an absence of modulation.  
These observations suggest than any diurnal sidereal modulation of conventional origin in the signal DM dataset is negligibly small compared to the statistical uncertainty of the signal dataset.  
This time-modulation cross-check was also performed for otherwise signal-like events with PSD parameters within 2$\sigma$ of the the electron recoil band center shown in Fig.~\ref{fig:PSDParameters}.  
This dataset also yielded an event rate ratio, 0.985$\pm$0.002,  consistent with a lack of unexpected modulation.  

\section{Dark Matter Search Results}\label{sec:analysis}

PROSPECT records a substantial rate of DM candidate events, most of which are presumably due to Standard Model backgrounds. However, we can still exclude strongly interacting DM by searching for the expected diurnal sidereal modulation of the signal, an effect that has been explored--but not yet used to set limits--in Refs.~\cite{DiGregorio:1993er,Has97,Kouvaris:2014lpa,Kouvaris:2015xga,Ge:2020yuf,Bla21}. Because of the large DM cross sections we consider, DM arriving from below the horizon is blocked by the Earth, and of the flux arriving from above the horizon, the further from vertical the initial trajectory is, the more the flux will be attenuated. As the Earth rotates, so does the vertical direction at PROSPECT's location, and the Galactic Center varies from being about 25 degrees above the horizon to being completely blocked by the Earth. As the flux of upscattered DM reaching PROSPECT scales with the line-of-sight integrated DM density times CR density, which is highest in the direction of the Galactic Center, this rotation produces a characteristic modulation over the course of a sidereal day.

\begin{figure}[hptb!]
	    \centering
   	    \includegraphics[width=\columnwidth]{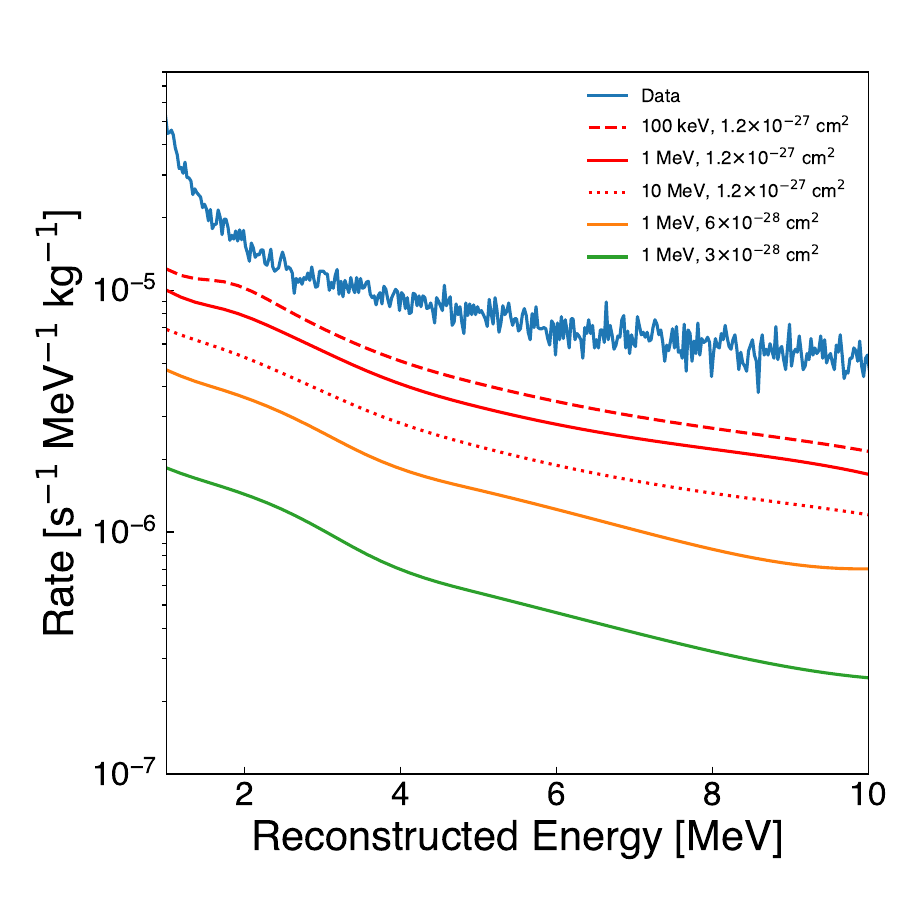}
	    \caption{Reconstructed energy spectrum of PROSPECT signal events over all time, compared with predicted DM-induced event spectra for several values of $m_{\chi}$ and $\sigma_{\chi N}$ during the time period from 00:00 to 01:00 GMST.}
	    \label{fig:EventSpectra}
	\end{figure}

When a DM particle collides with a proton in the detector, the  reconstructed energy in PROSPECT is related to the recoil energy of the proton by an energy-dependent quenching factor, as discussed in Sec.~\ref{sec:detector}.  
For modeling reconstructed energies of the DM-induced proton recoil signal, we use the \enquote{Birk9} fit from Ref.~\cite{NORSWORTHY201720}.  
Figure \ref{fig:EventSpectra} shows the predicted DM-induced event rate in PROSPECT for a few example DM masses and cross sections representing diverse regions of the parameter space that PROSPECT is sensitive to.   
Spectra are shown for times integrated from 00:00 to 01:00 GMST, close to when the DM signal is predicted to be strongest.  
Above a reconstructed energy of 1~MeV, the predicted shape of the DM signal's reconstructed energy spectrum exhibits a gradual downward trend with increasing energy, similar to the observed spectrum. This spectrum is determined just by the incoming DM spectrum (after attenuation) and the kinematics of the collisions, analogous to the computation of the DM spectrum in Eq.~\ref{DMdist2}.  
While the normalization of the predicted DM signal varies substantially across the DM phase space region of interest, its spectrum shape appears to be relatively consistent across this space.  
Due to the predicted spectrum's relative flatness and lack of finer-scale features, predicted DM signal event counts in the 1.5-10~MeV reconstructed energy range of interest are largely insensitive to other aspects of detector response, such as PROSPECT's photo-statistics or geometry-dependent energy resolution contributions.  
As uncertainties in the assumed proton quenching model are correlated across all time bins in the analysis, they also play a negligible role in defining the exclusion limits of this analysis.  
The latter point was verified by checking for consistency in DM exclusion contours between analyses incorporating each of the different proton quenching models referenced in this paper~\cite{NORSWORTHY201720,quench1,quench2,quench3}.  

\begin{figure}[phtb!]%
    \centering
    \includegraphics[width=\columnwidth]{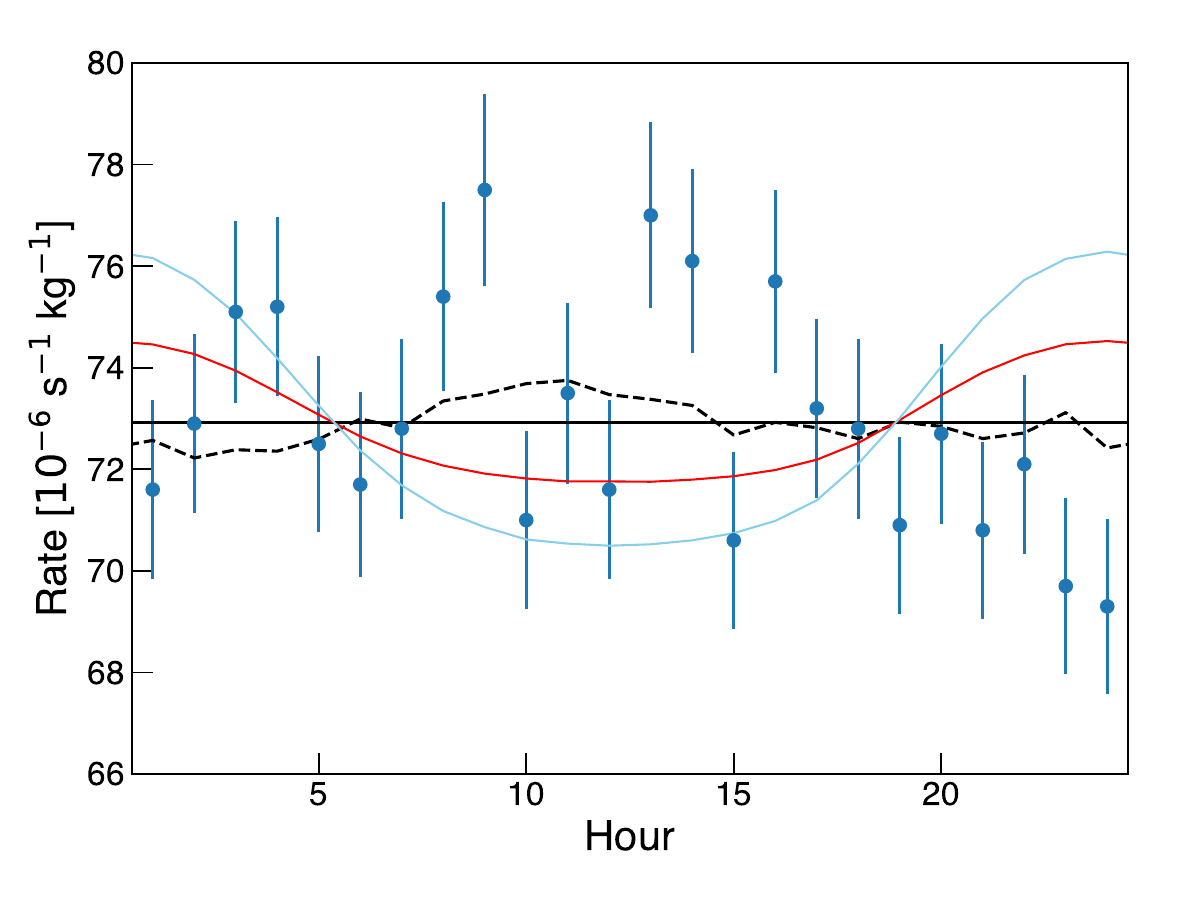}
    \caption{\label{fig:asymmetry} PROSPECT selected signal event rates plotted by time of sidereal day in 1~h bins. Data point error bars represent 1$\sigma$ statistical fluctuations. Prediction curves are for a time-independent Standard Model background (black, solid); a time-independent background plus a time-varying DM flux for $m_{\chi} = 1$ MeV and $\sigma_{\chi N} = 3\times10^{-28}$ cm$^2$ (red, solid); and a similar DM fit with $m_{\chi} = 1$ MeV and $\sigma_{\chi N} = 5\times10^{-28}$ cm$^2$ (light blue, solid). The dashed black curve shows hourly rate correction factors applied to all predictions to account for expected variations in cosmogenic backgrounds.  Hour 1 corresponds to time between 0:00 and 1:00 GMST.}
\end{figure}

As illustrated in Fig.~\ref{fig:EventSpectra}, the DM-induced event rate is smaller than the observed rate of signal-like events in PROSPECT in much of the parameter space of interest.  
This necessitates a time-binned analysis, as mentioned above and in Sec.~\ref{sec:crdm}.  
Fig.~\ref{fig:asymmetry} shows the signal event rate reported by PROSPECT in the 1.5-10~MeV energy range, plotted by time of sidereal day in 24 bins one sidereal hour in width.  
We note that, due to the shortness of the dataset relative to a solar year and relatively small frequency offset between sidereal and solar time (4~min per solar day), the phase offset between solar and sidereal time in this analysis is roughly consistent.  
For example, the sidereal time of highest expected DM signal, 23:00 to 00:00 GMST, corresponds to roughly 07:24 to 08:24 and 6:26 to 7:26 Eastern Daylight Time at the beginning and end of the dataset, respectively.  
Qualitatively, the data show no obvious indications of diurnal sidereal modulation.  


To quantitatively determine which regions of DM parameter space produce signal-like event rate modulations inconsistent with the observed data as plotted in Fig.~\ref{fig:asymmetry}, we define the following test statistic: 
\begin{equation}\label{chisqr}
\Delta\chi^2 = \chi^2_{DM} - \chi^2_{const}.
\end{equation}
In this test statistic, $\chi^2_{const}$ is defined as a one-parameter flat-line $\chi^2$ fit to the data as binned in~Fig.~\ref{fig:asymmetry}; $\chi^2_{DM}$ is defined by a predicted modulating DM contribution specific to each ($m_{\chi} $,$\sigma_{\chi N}$) phase space point, added to a fitted flat-line background contribution.  
Rates for $\chi^2_{const}$ and $\chi^2_{DM}$ predictions are corrected to account for expected percent-level variations in cosmogenically produced signal-like backgrounds, as described in Sec.~\ref{sec:detector}; these corrections are illustrated in Fig.~\ref{fig:asymmetry}.  
This figure also depicts signals for $\chi^2_{const}$ and for $\chi^2_{DM}$ for two test points in dark matter phase space prior to the application of rate correction factors.  
The black line depicts the best constant fit (minimum $\chi^2_{const}$/d.o.f. = 35.1/23) with respect to the data, which corresponds to the expected signal from a Standard Model background free from modulating DM effects.  
The red and blue curves represent $\chi^2_{DM}$ for test points ($m_{\chi},\sigma_{\chi N}$) = (1~MeV, 3$\times10^{-28}$ cm$^2$) and (1~MeV, 5$\times10^{-28}$ cm$^2$), respectively.  
Minimizing over the remaining parameter gives best-fit background rate contributions of 69.3~$\times$10$^{-6}$~s$^{-1}$\,kg$^{-1}$ and 65.1~$\times$10$^{-6}$~s$^{-1}$\,kg$^{-1}$ for these test points, respectively. Comparing this time-independent background rate with the time dependence of the total event rate apparent in Fig.~\ref{fig:asymmetry}, one can see that the DM event rate roughly doubles, from minimum to maximum, over the course of a sidereal day. For example, for a cross section of 5$\times10^{-28}$ cm$^2$, subtracting the background from the total event rate yields a DM event rate that varies from roughly 5.1~$\times$10$^{-6}$~s$^{-1}$\,kg$^{-1}$ to 10.9~$\times$10$^{-6}$~s$^{-1}$\,kg$^{-1}$. Both DM-including test points provide relatively poor fits to the observed data, with minimum $\chi^2_{DM}$/d.o.f. of 60.1/23 and 103.1/23 for the red and blue curves, respectively.  


\begin{figure}[phtb!]%
    \centering
    \includegraphics[width=\columnwidth]{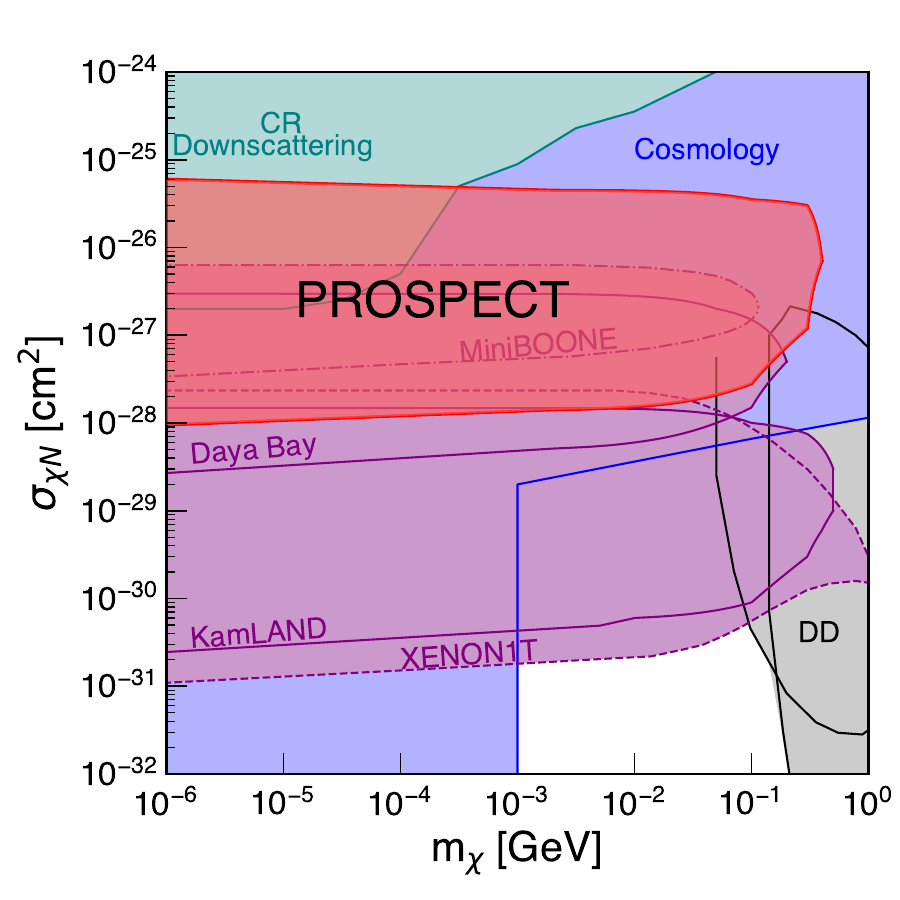}
    \caption{\label{fig:results} 95\% CL exclusion region from PROSPECT data (red) compared to other limits on CR-upscattered DM from Refs. \cite{Cap19b,Bri18} (purple), CR downscattering \cite{Cap19} (teal), cosmology \cite{Krn19,Nad19,Nad20} (blue), and direct detection \cite{Col18,Abd19} (gray).}
\end{figure}

By performing similar tests at an array of ($m_{\chi},\sigma_{\chi N}$) phase space points, we have determined excluded regions of dark matter parameter space using PROSPECT data as shown in Fig.~\ref{fig:results}.  
To assign exclusion confidence intervals, we use the Gaussian CL$_s$ method~\cite{cls}, which is useful in the context of performing searches for new physics in a continuous parameter space with large sample sizes.  
The CL$_{s}$ value determined by PROSPECT's dataset $x$ (the data points in Fig.~\ref{fig:asymmetry}) for each phase space point is defined as
\begin{equation}\label{eq:cls_detail}
\textrm{CL}_{s}(x) =  
\frac{1+\textrm{Erf}\large(\frac{\overline{\Delta T_{1}}- \Delta T(x)}{\sqrt{8|\overline{\Delta T_{1}}|}}\large)}{1+\textrm{Erf}\large(\frac{\overline{\Delta T_{0}}- \Delta T(x)}{\sqrt{8|\overline{\Delta T_{0}}|}}\large)}.
\end{equation}
Here, $\Delta T(x)$ is $\Delta\chi^2(x)$ (Eq.~\ref{chisqr}) for PROSPECT's measured dataset, $\overline{\Delta T_{0}}$ is $\chi^2_{DM}(x_{H0})$, where $x_{H0}$ denotes the Asimov dataset following the modulation-free hypothesis, and $\overline{\Delta T_{1}}$ is $-\chi^2_{const}(x_{H1})$, where $x_{H1}$ denotes the Asimov dataset following the dark matter signal for the phase space point in question. Phase space points with CL$_s$ values lower than 0.05 are disfavored by the data at the 95\% confidence level.  
The darkly shaded PROSPECT exclusion region in Fig.~\ref{fig:results} covers space previously unaddressed by other terrestrial particle physics experiments. The exclusion region is similar in size to the most optimistic projection derived by Ref.~\cite{Cap19b}, which assumed both significantly reduced background and improved background modeling. Taking advantage of the daily modulation of the DM signal was crucial to reach this sensitivity. 

The exclusion's lower limit is defined by the low fraction of incident dark matter flux interacting within the detector, while its upper limit is defined by attenuation of the dark matter flux prior to reaching the active detector region.  
Due to the relatively similar spectrum shapes between background and signal, negligible additional exclusion power is provided through a finer binning of the statistical analysis in energy.  
Expanded energy ranges for the analysis also offer limited improvement in exclusion power due to increased background rates at lower energies and low statistics at higher energies.

The strength of the DM exclusion does depend on assumptions about the half-height of the galactic CR halo.  
We adopted a commonly used value of 4 kpc, but estimates range from roughly 3-7 kpc (see Ref.~\cite{Str07} and references therein).  
Adjustments of halo half-height within this range result in reduction or expansion of limits at low cross section by less than a factor of 2; exclusions at high cross section are largely unaffected. Similarly, the daily modulation of DM events depends on the assumed DM profile. We tested the robustness of our results by comparing two alternative DM halos: a more concentrated NFW profile with a scale radius of 10 kpc, and an extremely cored model, which follows an NFW profile at large radii but has constant density within about 8 kpc. The more concentrated NFW profile increases the amplitude of the daily modulation, while the cored profile decreases it, but in both cases only by tens of percent. This produces an O(10\%) variation in the strength of our limit at low cross sections. 

We have also considered whether the observed hourly signal-like rates in Fig.~\ref{fig:asymmetry} are consistent with a sinusoidal modulation beyond that allowed by the boosted dark matter signal of central concern in this paper.  
A best-fit modulation was found with an amplitude 1.46\% of the total signal rate and a phase approximately 12 hours behind the DM signal.  
This fit, which has the phase and amplitude of the sinusoid as free parameters and thus includes 2 fewer degrees of freedom, provided a $\chi^2$ of 30.72, 4.33 below that of the $\chi^2_{const}$ fit described above.  
A frequentist approach was then employed to determine the strength of this apparent preference towards a modulated signal.  
Similar $\Delta \chi^2$ values were calculated for 10$^3$ simulated modulation-free PROSPECT datasets with statistical fluctuations matching those of the observed dataset.  
The $\Delta \chi^2$ value of 4.33 derived from the observed data is lower than 22.2\% of values from simulated datasets, indicating that the observed data is consistent with a lack of daily modulation. 


\section{Conclusions}\label{sec:conclusions}

In summary, we have used a dedicated analysis of 14.6 solar days of the PROSPECT neutrino experiment's reactor-off data to provide new bounds on the nature of dark matter.  
This search is enabled by PROSPECT's unique experimental configuration, which combines on-surface detector deployment with powerful particle discrimination and event topology reconstruction capabilities.  
After applying analysis cuts designed to select isolated proton recoil signatures within 440~kg of target liquid scintillator, we have identified 37522 candidate interactions of energetic dark matter upscattered by cosmic rays.  
As signal detection rates do not exhibit any statistically significant degree of diurnal sidereal modulation, as would be produced by strongly interacting dark matter originating in the galactic halo, we are able to exclude the existence of dark matter over a broad range of sub-GeV parameter space.  
This new constraint addresses phase space regions beyond those accessible in traditional or low-threshold direct-detection experiments, and reaches cross sections about an order of magnitude larger than those previously probed in other studies of cosmic ray upscattered dark matter.  
Our limit is complementary to existing constraints from cosmology and structure formation: while these other limits are indirect constraints based on scattering in the early Universe, our result is a direct-detection limit based on scattering in the present day.  
In the future, longer data collection times and improved background rejection could extend sensitivity substantially at low cross section, but only modestly at high cross section, where useful signatures are sharply cut off by extreme atmospheric attenuation of the incoming DM flux.  Mild improvements in sensitivity at high cross section may also be achieved through redeployment at high ($>$km) elevations.

\section*{Acknowledgements}\label{sec:ack}

We are grateful for helpful discussions with John Beacom in the preparation of this analysis and review of this manuscript.  

This material is based upon work supported by the following sources: U.S. Department of Energy (DOE) Office of Science, Office of High Energy Physics under Awards No. DE-SC0016357 and DE-SC0017660 to Yale University, under Award No. DE-SC0017815 to Drexel University, under Award No. DE-SC0008347 to Illinois Institute of Technology, under Award No. DE-SC0016060 to Temple University, under Contract No. DE-SC0012704 to Brookhaven National Laboratory, and under Work Proposal Number SCW1504 to Lawrence Livermore National Laboratory. This work was performed under the auspices of the U.S. Department of Energy by Lawrence Livermore National Laboratory under Contract No. DE-AC52-07NA27344 and by Oak Ridge National Laboratory under Contract No. DE-AC05-00OR22725. Additional funding for the experiment was provided by the Heising-Simons Foundation under Award No. \#2016-117 to Yale University. 
J.G. is supported through the NSF Graduate Research Fellowship Program and A.C. performed work under appointment to the Nuclear Nonproliferation International Safeguards Fellowship Program sponsored by the National Nuclear Security Administration’s Office of International Nuclear Safeguards (NA-241). This work was also supported by the Canada  First  Research  Excellence  Fund  (CFREF), and the Natural Sciences and Engineering Research Council of Canada (NSERC) Discovery  program under grant \#RGPIN-418579, and Province of Ontario.  
C. C. is supported through NSF Grant No. PHY-2012955.  

We further acknowledge support from Yale University, the Illinois Institute of Technology, Temple University, Brookhaven National Laboratory, the Lawrence Livermore National Laboratory LDRD program, the National Institute of Standards and Technology, and Oak Ridge National Laboratory. We gratefully acknowledge the support and hospitality of the High Flux Isotope Reactor and Oak Ridge National Laboratory, managed by UT-Battelle for the U.S. Department of Energy.

\bibliography{PROSPECT}{}

\pagebreak

\end{document}